\documentclass[journal]{IEEEtai}

\usepackage[colorlinks,urlcolor=blue,linkcolor=blue,citecolor=blue]{hyperref}

\usepackage{color,array}

\usepackage{graphicx}
\usepackage{tabularx}
\usepackage{longtable}
\usepackage{bbding}
\usepackage{array}
\usepackage{supertabular}
\usepackage{booktabs}
\usepackage{epstopdf}

\begin{document}

\title{Intellectual Property Protection for Deep Learning Models: Taxonomy, Methods, Attacks, and Evaluations}

\author{Mingfu~Xue,~Yushu~Zhang,~Jian~Wang,~and~Weiqiang~Liu
\thanks{M. Xue, Y. Zhang and J. Wang are with the College of Computer Science and Technology, Nanjing University of Aeronautics and Astronautics, Nanjing, 211106, China (e-mail: mingfu.xue@nuaa.edu.cn; yushu@nuaa.edu.cn; wangjian@nuaa.edu.cn).}
\thanks{W. Liu is with the College of Electronic and Information Engineering, Nanjing University of Aeronautics and Astronautics, Nanjing, 211106, China (e-mail: liuweiqiang@nuaa.edu.cn).}
}

\markboth{}
{Xue \MakeLowercase{\textit{et al.}}: Intellectual Property Protection for Deep Learning Models: Taxonomy, Methods, Attacks, and Evaluations}

\maketitle

\begin{abstract}
The training and creation of deep learning model is usually costly, thus it can be regarded as an intellectual property (IP) of the model creator.
However, malicious users who obtain high-performance models may illegally copy, redistribute, or abuse the models without permission. To deal with such security threats, a few deep neural networks (DNN) IP protection methods have been proposed in recent years.
This paper attempts to provide a review of the existing DNN IP protection works and also an outlook.
First, we propose the first taxonomy for DNN IP protection methods in terms of six attributes: scenario, mechanism, capacity, type, function, and target models.
Then, we present a survey on existing DNN IP protection works in terms of the above six attributes, especially focusing on the challenges these methods face, whether these methods can provide proactive protection, and their resistances to different levels of attacks.
After that, we analyze the potential attacks on DNN IP protection methods from the aspects of model modifications, evasion attacks, and active attacks.
Besides, a systematic evaluation method for DNN IP protection methods with respect to basic functional metrics, attack-resistance metrics, and customized metrics for different application scenarios is given.
Lastly, future research opportunities and challenges on DNN IP protection are presented.

\end{abstract}

\begin{IEEEImpStatement}
Building a high-performance deep neural networks (DNN) model is costly, thus the trained DNN model is oftentimes regarded as an intellectual property (IP) of the model creator. The IP infringement of the DNN model has caused serious concerns in recent years. This paper presents a survey on DNN IP protection methods. First, we propose the first taxonomy for DNN IP protection methods in terms of six attributes. Then, we summarize the existing DNN IP protection works with a focus on the challenges they face as well as their ability to provide proactive protection and resistance to different levels of attacks. After that, the potential attacks on existing methods are analyzed, and a systematic evaluation method for DNN IP protection methods is given. Finally, future research opportunities and challenges are prospected. This paper can hopefully provide a reference for the taxonomy, comparison, evaluation and development of DNN IP protection methods.
\end{IEEEImpStatement}


\section{Introduction}
Deep learning (DL) techniques, especially deep neural networks (DNN), have been widely applied in many tasks, e.g., image classification, object detection, voice recognition, natural language processing, driverless cars \cite{xue2020machine}.
The training process of the DNN model requires massive training data, expensive hardware resources, and often takes weeks or even months, which requires high cost. This leads to that ordinary users are difficult to train high-accuracy DNN models, and deep learning models are often provided by large companies.
This increasingly popular business model is called Machine Learning as a Service (MLaaS). Deep learning models have high business value, thus can be considered as an intellectual property (IP) \cite{UchidaNSS17,NagaiUSS18,RouhaniCK19,abs-1811-03713} of the model creators, which need to be protected. With the widespread applications of DNN models, the IP infringement of DNN is an emerging problem \cite{UchidaNSS17,NagaiUSS18,RouhaniCK19,abs-1811-03713}, which attracted serious concerns in recent years. DNN IP protection is a frontier research field, which is still in its infancy.

\begin{figure*}[!htbp]
  \centering
  \includegraphics[width=0.7\linewidth]{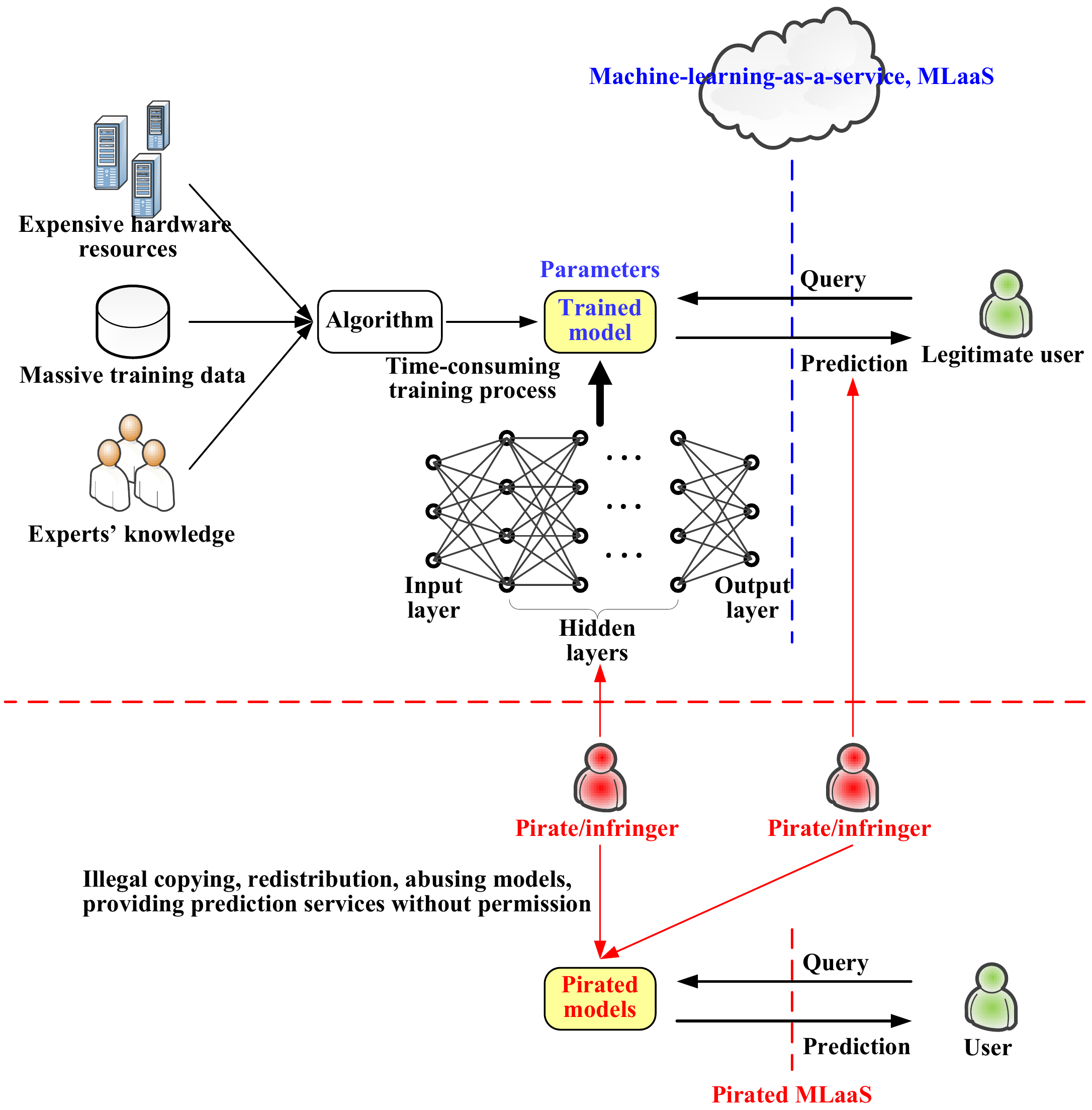}
  \caption{Overview of piracy on deep learning models}
    \label{fig:overview}
\end{figure*}

DNNs are deployed in white-box scenarios that expose internal models, or in black-box scenarios where only the model's output is known.
In black-box scenarios, prediction application programming interfaces (APIs) are often provided to users as paid services. However, malicious users who obtain high-performance models may illegally copy, redistribute, abuse the models, or use the models to provide prediction services without permission, as shown in Figure \ref{fig:overview}. In addition, some model users may inadvertently disclose the parameters and architecture of the model to the public. In order to deal with these security threats, a method capable of verifying model ownership from the outside is needed \cite{UchidaNSS17,NagaiUSS18,abs-1811-03713}. A promising solution is to embed watermarks in the models so that the copyright owners can verify the ownership of the models from the outside \cite{RouhaniCK19, NambaS19}.

Since 2017, a few DNN IP protection works have been proposed. However, existing DNN IP protection works face the following challenges: (i) Most of them are passive verification methods afterwards, which cannot actively prevent the occurrence of piracy. (ii) Most of the existing works only focus on the copyright verification of the model, but do not authenticate and manage the users' unique identities, which cannot provide copyright management function for commercial applications. Besides, this can also lead to attacks launched by dishonest users, such as user collusion attacks. (iii) Most of the existing works only evaluated the attack-resistance to model modifications, and rarely consider the robustness and security when the pirates take active attacks. As a result, attackers can launch some active and strong attacks to subvert existing DNN IP protection methods. (iv) As an frontier research field, DNN IP protection is still in its infancy, which lacks of systematic evaluation method and evaluation metrics. This will restrict the applications of DNN models.

This paper attempts to provide a review of the existing DNN IP protection works and also an outlook.
We have published a previous conference version \cite{glvlsi} of this work in ACM GLSVLSI as an invited paper. This paper is the extended version of our previous work \cite{glvlsi}, and the new materials and contributions of this work are as follows: (i) A detailed survey on existing DNN IP protection methods in terms of the six attributes are presented in Section \ref{sec3}; (ii) Discussion about the watermarking techniques in the multimedia field cannot be directly applied to DNN watermarking; (iii) Explained in detail the three levels of attacks, the various attack methods, and related works. Besides, the attack-resistance of existing works are also discussed; (iv) The meaning of all the evaluation metrics is given, as shown in Table \ref{tab:3}; (v) A detailed discussion on challenges and future works are presented.

The contributions of this paper are as follows:
\begin{enumerate}
\item{\textbf{The first taxonomy for DNN IP protection methods is proposed.} For the first time, we propose a taxonomy for DNN IP protection methods in terms of the following six attributes: \textit{scenario}, \textit{mechanism}, \textit{capacity}, \textit{type}, \textit{function}, and \textit{target models}. Such a theoretical taxonomy can facilitate the analysis and comparison of existing methods and the development of future methods.}

\item{\textbf{The survey on existing DNN IP protection methods in terms of the above six attributes are presented}, especially focusing on the challenges these methods face, whether these methods can provide proactive protection, and their resistances to different levels of attacks.}

\item{\textbf{Analysis on attacks}. We divide the possible attacks on DNN IP protection methods into three levels (from weak to strong): (i) model modifications; (ii) evasion attacks and removal attacks (passive attacks); (iii) active attacks. Compared to existing works, one of the contributions of this paper is to analyze the attack-resistance of the DNN IP protection methods in the case of active attacks launched by the adversary.}

\item{\textbf{Systematic evaluation suggestions for DNN IP protection methods are presented}. Most of the evaluations in the existing DNN IP protection works only focus on the functional metrics of the DNN watermark. We suggest to build the evaluation method for DNN IP protection methods from the following aspects: (i) systematic evaluation method; (ii) basic functional metrics and attack-driven metrics; (iii) evaluating the DNN IP protection methods when the attackers take different levels of attacks.}

\item{\textbf{Challenges and future works}. We discuss the challenges faced by the state-of-the-art DNN IP protection methods and present the insights on future works.}
\end{enumerate}

This paper is organized as follows. The proposed taxonomy for DNN IP protection methods is presented in Section \ref{sec2}. The survey on existing DNN IP protection works is presented in Section \ref{sec3}. Three levels of attacks and the attack-resistance of existing works are discussed in Section \ref{sec4}. Evaluation suggestions for DNN IP protection methods are presented in Section \ref{sec5}. The challenges and future works on DNN IP protection are discussed in Section \ref{sec6}. This paper is concluded in Section \ref{sec7}.

\section{Taxonomy}
\label{sec2}

\begin{figure}[!htbp]
  \centering
  \includegraphics[width=1\linewidth]{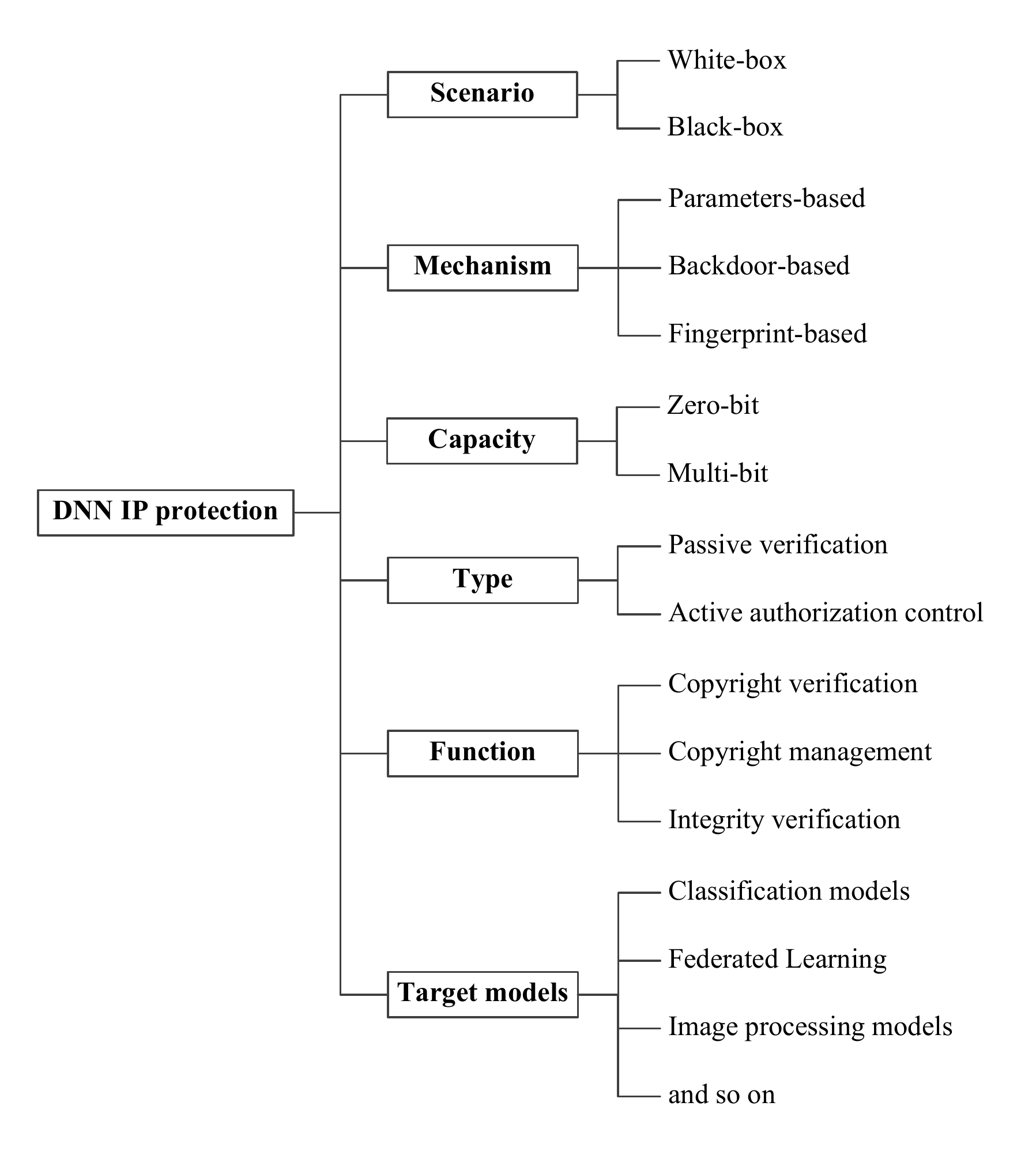}
  \caption{The proposed taxonomy for DNN IP protection methods}
    \label{fig:taxonomy}
\end{figure}

To date, a few DNN IP protection works have been proposed. However, there is still no systematic taxonomies.
In this paper, we propose a taxonomy for DNN IP protection methods in terms of the following six attributes, as shown in Figure \ref{fig:taxonomy}:

1) \textbf{Scenario}. In a \textit{white-box} scenario, the internal parameters of the model to be verified are publicly available.
However, in practice, the DNN model is often deployed as an online service, and only prediction and confidence are provided through an API, which is called the \textit{black-box} scenario.

2) \textbf{Mechanism}. The implementation mechanism of the DNN IP protection methods can be divided into the following categories: \textit{parameter-based} (embedding a watermark in the weights/parameters of the model); \textit{backdoor-based} (using DNN backdoor as the watermark of the model); \textit{fingerprint-based} (using the data distribution of the model's prediction to specific inputs, e.g., adversarial examples, as the fingerprint of the model).

3) \textbf{Capacity}. Capacity represents the amount of information that the watermark method can embed. If the verification scheme only focuses on the presence of the watermark, it is called a \textit{zero-bit} scheme. If the scheme can perform multi-bit string verification instead of verifying one bit of information \cite{abs-1904-00344}, it is called a \textit{multi-bit} scheme.

4) \textbf{Type}. If the method passively verifies the copyright of the model after the piracy occurs, it is called \textit{passive verification}. If the scheme can actively control the useage of the model through authorization control to prevent piracy, it is called \textit{active authorization control}.

5) \textbf{Function}. The function of DNN IP protection works can be divided into three catenaries: (i) \textit{Copyright verification}, which means verifying the ownership of the model using robust watermarks. This is the most common situation. (ii) \textit{Copyright management}, which means managing users' identities and implementing authorization control. (iii) \textit{Integrity verification}, which means verifying the integrity of the model through fragile/reversible watermarks.

6) \textbf{Target models}. Most DNN IP protection methods targeted at classification tasks. Recently, few works targeted at Federated Learning (distributed scenarios), while there are also few IP protection works that targeted at image processing tasks. Depending on various application scenarios, IP protection methods for different tasks are needed.

\section{Survey on DNN IP Protection Works} \label{sec3}
In the multimedia field, digital watermarking techniques have been extensively used to protect the copyright of multimedia data (such as images, videos, etc.). However, IP protection in the context of deep learning is still in its infancy. Unlike embedding the digital watermarks into the multimedia content, it is necessary to design new methods to embed watermarks into the DNN model, while existing digital watermarking techniques cannot be applied directly \cite{ZhangGJWSHM18}. Existing digital watermarking algorithms require direct access to multimedia content to extract the watermark. However, unlike the multimedia data, the DNN model has complex structure and massive parameters. Usually only the APIs of the DNN models are available for watermark extraction and ownership verification \cite{ZhangGJWSHM18}. Therefore, existing digital watermarking techniques are not suitable for the scenarios of DNN.

In addition, developing a practical DNN IP protection technique is extremely challenging due to the following reasons \cite{abs-1904-00344, ChenRFZK19, ChenFRZK19}:
(i) A public watermarking algorithm is needed, which can be used to credibly verify the ownership of the model multiple times through API only;
(ii) Watermark embedding should not cause the performance of the model to be degraded;
(iii) The embedded watermark should be able to generate a high detection rate, and produce a minimum false alarm rate to avoid mistakenly accusing innocent users of misusing/stealing the models;
(iv) Users can fine-tune or pruning the model to modify its parameters;
(v) The watermark should be robust to potential attacks by malicious users.

\begin{figure}[!htbp]
  \centering
  \includegraphics[width=1\linewidth]{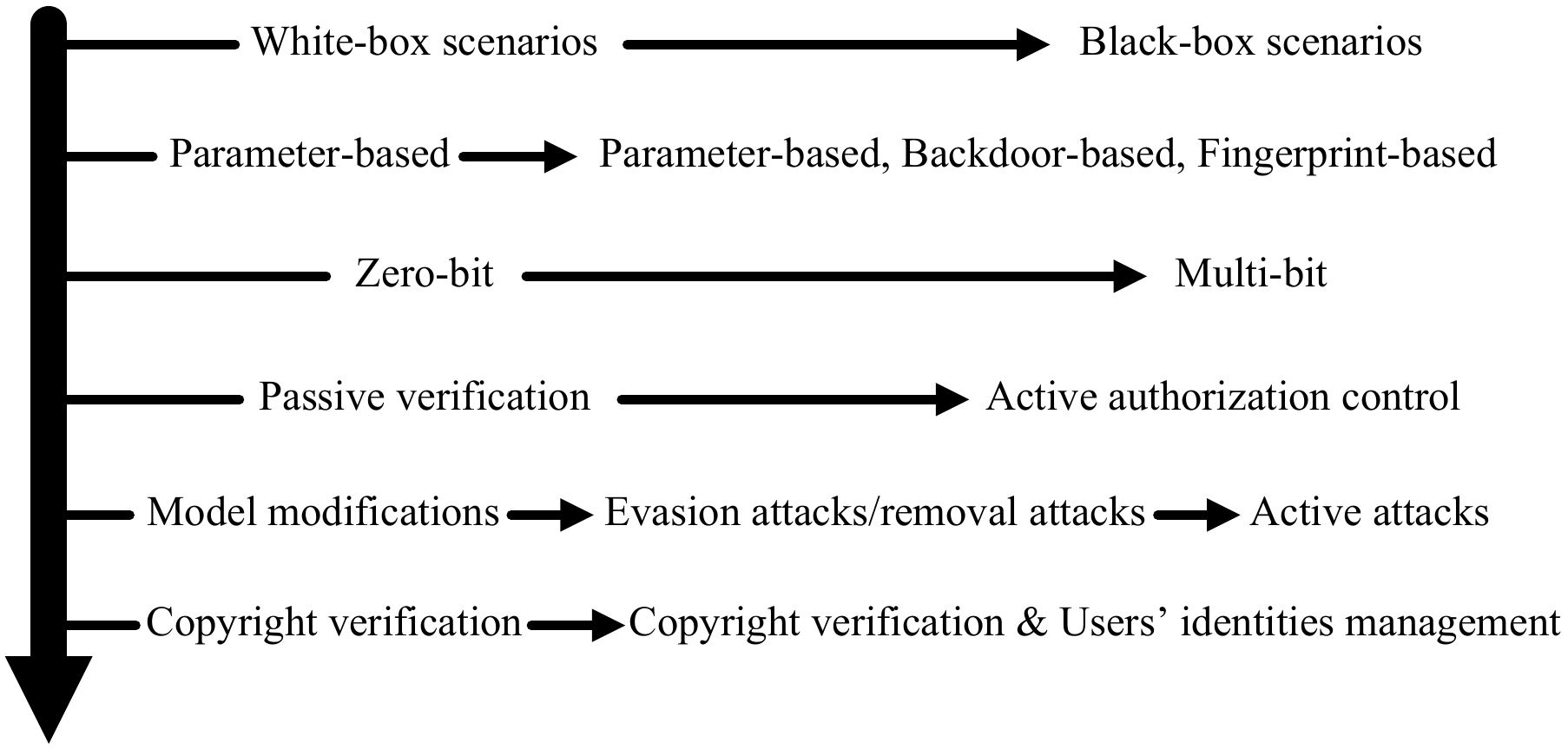}
  \caption{The pipeline of DNN IP protection works}
    \label{fig:pipeline}
\end{figure}

The existing DNN IP protection works are reviewed in the following sections with respect to the taxonomy proposed in Section \ref{sec2}.
The pipeline of DNN IP protection works is shown in Figure \ref{fig:pipeline}.

\subsection{Scenarios}

\subsubsection{White-box Scenarios.}
Uchida \textit{et al.} \cite{UchidaNSS17,NagaiUSS18} proposed the first DNN copyright protection method. They trained the model with additional regularization loss to embed the watermark in the weight of the middle layer. In the verification stage, the watermark can be extracted from the weight of the marking layer \cite{UchidaNSS17}.
Wang and Kerschbaum \cite{wang2019riga} propose an adversarial training based watermarking approach for white-box scenarios, named RIGA, with the purpose of not affecting the accuracy and be robust. They construct a structure similar to Generative Adversarial Networks (GAN), where the training/watermarking process of the model and the watermark detector are two competing parties \cite{wang2019riga}.
These methods work in the white-box scenario, where the internal parameters of the model to be verified are publicly available.
However, in practice, the pirate often deploys the pirated DNN model as an online service which only outputs prediction and confidence through a remote API. The verifier cannot obtain the internal information of the suspicious model, i.e., the black-box scenario. Most of the deep learning APIs deployed in cloud servers are in black-box scenarios.

\subsubsection{Black-box Works}
Most of the existing DNN IP protection works focus on the black-box scenarios. The working mechanism is that, a watermark is embedded in the model, and the watermark information can only be extracted from the remote model by interacting with the model through a remote API \cite{abs-1811-03713}.

Rouhani \textit{et al.} \cite{RouhaniCK19} proposed a general watermarking method (named DeepSigns), which is applicable to both black-box and white-box scenarios.
DeepSigns embeds a $N$-bit string (the owner's signature/watermark) into the probability density function of the activation set of each layer. The embedded watermark can be triggered by a corresponding set of input to remotely verify the copyright of deep neural networks \cite{RouhaniCK19}.

\subsection{Mechanism}
\subsubsection{Parameters-based Watermarking}
In many existing DNN IP protection works, the watermark is embedded in the weights/parameters of the models. Deep learning model has a large number of parameters and can ``memorize'' information far beyond the main task, which can be used for watermark embedding \cite{song2017machine}.

However, in the parameters-based watermarking methods, the weights change greatly. The existence of the watermark can be detected by analyzing the change of the weights. To this end, Kuribayashi \textit{et al.} \cite{DeepWatermark} propose a quantifiable watermark embedding approach on the weights of fully connected layers. By changing the parameters during training, the influence of the watermark can be quantified, so as to ensure that the change caused by the embedded watermark is small \cite{DeepWatermark}.

Feng and Zhang \cite{FengZ20} propose a watermarking approach with compensation. They select random positions for watermark embedding. After watermark embedding, they use fine-tuning with compensation to ensure that the watermark embedding will not reduce the accuracy of the model \cite{FengZ20}.

In addition to embedding a watermark in the model, the output image can also be watermarked. Wu \textit{et al.} \cite{WatermarkedImages} propose an approach, where the images output from a watermarked DNN also contain a watermark. They train the host DNN and the watermark extraction network together, and use a combined loss function, so that the host DNN can complete the task and output watermarked images \cite{WatermarkedImages}. In addition to identifying the copyright of the model, the method can also be used to determine whether an image is output by the specific model.

Some researchers also propose encryption based approaches to protect the model. Pyone \textit{et al.} \cite{pyone2020training} encrypt the training images through block-wise pixel shuffling based on a key. The model is trained on these preprocessed encrypted images. The user can only obtain the normal model performance by inputting the encrypted image \cite{pyone2020training}.
Gomez \textit{et al.} \cite{GomezIMD18,GomezWMD19} use fully homomorphic encryption to protect the IP, the input data, and the inference of the neural networks.
The efficiency of the fully homomorphic encryption based scheme is low and the method may affect the accuracy of the model.
Lin \textit{et al.} \cite{Chaotic} propose a scheme based on chaotic weights for DNN IP protection. They exchange the positions of weights to make them chaotic without affecting the accuracy of the model.
However, these encryption based approaches may affect the model performance or introduce high overhead.
Xue \textit{et al.} \cite{AdvParams} propose an adversarial perturbation based parameter encryption scheme to protect the intellectual property of DNN. The method only needs to disturb a very small number of parameters with low overhead, while can actively prevent the infringement in advance.

Sakazawa \textit{et al.} \cite{SakazawaMTY19} propose the accumulation and visual decoding of watermarks in the DNN model, and allowed the third-party verification by providing a subset of the decoded data.

\subsubsection{Backdoor-based Method}
The backdoor attack \cite{abs-1712-05526, xue2020TDSC} used to be an attack method on deep learning models where the attacker trains the model so that when a specific input is arrived, the model will output the specified label. Yossi \textit{et al.} \cite{AdiBCPK18} use a backdoor as the watermark key image, and use overparameterization of the neural network to implement the watermark scheme. The corresponding key label is randomly selected from all the classes except the true label and the original predicted label. The watermark is detected by comparing the accuracy of the watermark trigger set with a threshold \cite{AdiBCPK18}. In addition, they use a commitment scheme to build a publicly verifiable protocol.
Guo and Potkonjak \cite{abs-1906-04411} proposed an evolutionary algorithm-based method to generate and optimize the trigger pattern for the backdoor-based watermark to reduce the false alarm rate. Zhang \textit{et al.} \cite{ZhangGJWSHM18} proposed three different watermark key generation methods using irrelevant images in another data set, training images superimposed with additional content, and random images, respectively. Then, they use the watermark key to fine-tune the pre-trained model. In the watermark detection stage, the owner sends the watermark key image to a remote DNN service, and sets a threshold for classification accuracy to make a Boolean decision \cite{ZhangGJWSHM18}.
Guo and Potkonjak \cite{GuoP18} proposed a black-box watermarking method for embedded applications.
In the framework, the signature of the copyright owner is embedded in the dataset to train a watermarked DNN.
When the DNN encounters any input with the embedded signature, it will run in a predefined temporary mode, thereby verifying the identity of the owner \cite{GuoP18}.
Li \textit{et al.} \cite{LiHZG19} proposed a DNN IP protection framework based on blind watermarking. The framework aims to generate key instances with a distribution similar to the original instances, and explicitly associate the model with the model creator's identity. The framework takes ordinary instances and specific labels as inputs, generates watermark key instances, and embed the watermark into the DNN model \cite{LiHZG19}.

Li \textit{et al.} \cite{li2019piracy} propose a watermarking embedding method that can only be inserted during the initial training of the model, named NULL embedding. A strong dependence will be established between the accuracy and the embedded parameters. Therefore, the attacker neither can remove the embedded watermark through incremental training or fine-tuning, nor can add a new watermark \cite{li2019piracy}.
The existing watermark embedding method has nothing to do with the main task, so it may be removed through model compression or transfer learning. Therefore, Jia \textit{et al.} \cite{Entangled} propose Entangled Watermarks. The watermark is entangled with the legal data of the model. As a result, if an attacker tries to remove the watermark, the performance of the model will decrease on the legal data \cite{Entangled}.

In the backdoor-based method, assigning a wrong label to the key sample may distort the decision boundary of the model thus will affect the model's performance \cite{ZhongZ0G020}. To address this problem, Zhong \textit{et al.} \cite{ZhongZ0G020} propose a black-box watermarking method in which a new label is assigned to the key sample. As a result, the model will learn the feature of the key sample while not distorting the decision boundary of the original model.
Zhang \textit{et al.} \cite{ZhangJWNC20} propose a automatic method based on chaos to label the backdoor samples.

Recently, many attacks against backdoor-based watermarking methods have been proposed. To this end, Zhu \textit{et al.} \cite{ZhuZST20} use one-way hash function for watermarking to resist forgery attacks. The trigger samples constitute a one-way chain with specified labels. In this way, it is impossible for an attacker to forge such trigger samples \cite{ZhuZST20}.

Embedding watermarks into the model allows IP owners to identify the ownership of the models after piracy. However, the function of the model may also be stolen through model extraction attacks \cite{TramerZJRR16,JuutiSMA19}. In this case, the adversary will use the predictions returned from the API of the model to train a substitute model. Existing watermarking methods are ineffective against model piracy through model extraction attacks because the substitute model is trained by the pirate (not the IP owner) \cite{abs-1906-00830}. Szyller \textit{et al.} \cite{abs-1906-00830} proposed a dynamic adversarial watermarking scheme for neural networks, named DAWN, to prevent model extraction attacks based IP stealing. DAWN is deployed in the prediction API of the model. It changes the prediction response to the customer by dynamically watermarking a small number of queries from the customer \cite{abs-1906-00830}. If the pirate uses the response of the query to train the substitute model, the watermarked query will be used as the trigger set, which can be used to verify the ownership of the substitute model later \cite{abs-1906-00830}.

\subsubsection{Fingerprint-based Method}
Some studies have demonstrated that the ``fingerprint'' of the model can be extracted for IP protection.
Merrer \textit{et al.} \cite{abs-1711-01894} proposed a watermarking algorithm using adversarial examples \cite{GoodfellowSS14, xue2021naturalAE} as the watermark key set. The method slightly adjusts the decision boundary of the model so that a specific set of queries can verify the watermark information.
To this end, they add perturbations to generate adversarial instances which are very close to the boundary of the model \cite{abs-1711-01894}.
In the watermark detection stage, the model is queried using the watermark key image. The remote model's responses to these adversarial inputs are compared with the marked model's responses. If the number of mismatches between the model's response and the watermark key label is less than the threshold, it is considered that there is a watermark in the remote model \cite{abs-1711-01894}.
Lukas \textit{et al.} \cite{ConferrableAE} use conferrable adversarial examples as the fingerprint of DNN model, which is resistant to distillation attacks. Specifically, they propose an ensemble method to generate the conferrable adversarial examples.
Zhao \textit{et al.} \cite{ZhaoHLMCH20} proposed a DNN fingerprint authentication method, named AFA, which exploits adversarial examples as the model's fingerprint for IP protection.
The small model modifications by the pirates cannot change or delete the fingerprint of the model. The transferability of the adversarial mark is used to determine whether the suspicious model is a pirated model \cite{ZhaoHLMCH20}.
The work judges whether it is a pirated model by verifying whether the response of the model to the adversarial example is consistent. However, two models may have the same response to adversarial examples, but they are not the same model.

During the watermark embedding, the training process of the classifier is modified, which may sacrifice the accuracy of the model \cite{abs-1910-12903}. In order not to affect the accuracy of the original classifier, Cao \textit{et al.} \cite{abs-1910-12903} propose an approach (named IPGuard) to protect the IP of the DNN classifier. They indicate that the DNN classifier can be uniquely represented by its classification boundary. IPGuard finds some data points near the classification boundary of the target classifier, and regards these data points and the labels predicted by the classifier as the fingerprint of the classifier \cite{abs-1910-12903}. For suspicious classifiers, the model owner queries its prediction API to obtain the labels of these data points. If the suspicious classifier and the target classifier predict the same labels for most fingerprint data points, the model owner verifies that the suspicious classifier is pirated from the target classifier. IPGuard does not modify the training process, so it will not cause accuracy drop to the target classifier \cite{abs-1910-12903}.

\subsection{Capacity}
Most of the current black-box watermarking methods are zero-bit watermarking, i.e., only verifying the presence of the watermarks \cite{abs-1904-00344}. Generally, they generate a set of watermark key pairs and strategically change the decision boundary of the target model. The watermark key image is used to query the model, then the corresponding accuracy is compared with the threshold to determine the existence of the watermark \cite{abs-1904-00344}. However, the watermark capacity is limited. Chen \textit{et al.} \cite{abs-1904-00344} proposed a multi-bit watermarking framework for the black-box scenarios, named BlackMarks. They demonstrate that it is feasible to use the prediction of the model to perform multi-bit string verification instead of verifying one bit of information. According to the owner's watermark signature, the scheme designs a set of key images and label pairs. Then, by using the generated watermark key set to fine-tune the model, the watermark is embedded in the behavior of the target DNN \cite{abs-1904-00344}. When extracting the watermark, the remote model will be queried using the watermark key image, and the owner's signature will be decoded from the corresponding prediction \cite{abs-1904-00344}.

\subsection{Type}
All the above black-box watermarking works are passive verification methods, i.e., the copyright of the model is passively verified after the piracy occurs.
Recently, few active authorization control methods \cite{AdvParams, DeepLock, DSN, ChenW18, FanNC19, Zhang00Z0Y20, TrustCom20, ActiveGuard, ACS} have been proposed.

Tang \textit{et al.} \cite{DSN} propose a DNN IP protection method based on serial number, which is implemented using a knowledge distillation approach. The teacher model is trained first, and then be distilled to a series of customer (student) models. Each customer model is assigned with a serial number, and the customer model can be used normally only if the correct serial number is entered \cite{DSN}. The serial number is considered as the watermark for copyright verification.

Chen and Wu \cite{ChenW18} proposed an access control framework for DNN so that only authorized users can use the model correctly. The framework enables DNN to maintain the function for authorized access, but does not work for unauthorized access or illegal use. A conversion module based on the adversarial examples is designed to provide authorized input. When an unauthorized user provides input to the model, it is disturbed by adversarial perturbations, which leads to poor performance \cite{ChenW18}. In contrast, authorized users can use the conversion module to pre-process the input to obtain high performance predictions \cite{ChenW18}. This method does not take into account the user's identity management (i.e., cannot distinguish different authorized users).

Fan \textit{et al.} \cite{FanNC19} indicate that the ambiguity attack poses a serious threat to the existing DNN watermarking methods. As a remedy, they proposed to embed specific passport layers into DNN, which can paralyze the function of the neural network for unauthorized use, or maintain its function under verified circumstances. Unless a valid passport is provided, the DNN model will not function properly, thereby preventing the illegal use of the model \cite{FanNC19}.
However, in this scheme, passport layers are added after each convolutional layer, which will introduce high overhead. Besides, the method is vulnerable to tampering attack and reverse-engineering attack.
Zhang \textit{et al.} \cite{Zhang00Z0Y20} also propose a passport-aware normalization paradigm for DNN IP protection. A new passport-aware branch is added, which is trained together with the model. The passport-aware branch will be discarded during test, and will only be re-added and work during ownership verification \cite{Zhang00Z0Y20}. Only by providing the correct passport can the performance of the model be maintained, otherwise the performance of the model will drop significantly \cite{Zhang00Z0Y20}.

The above authorization control methods do not take into account of the users' identities management, thus cannot distinguish different authorized users, and are vulnerable to attacks initiated by dishonest users, such as collusion attacks. Besides, due to the lack of copyright management function, these methods cannot satisfy the requirements of commercial DNN IP protection applications.
To this end, Xue \textit{et al.} \cite{TrustCom20} propose a copyright management framework for DNN. Based on the multi-trigger backdoor technique \cite{xue2020TDSC}, each legitimate user is assigned with few images with a small amount of backdoor signals. These images are used as the user's fingerprint and can trigger the backdoor with a certain medium probability, thus verifying whether the user is authorized or not \cite{TrustCom20}. Only the model owner has all the backdoor signals and can trigger the backdoor with a high probability, thereby being able to verify the ownership of the model \cite{TrustCom20}.
Sun \textit{et al.} \cite{ACS} propose an active DNN IP protection method, in which they train the model with additional class for ownership verification, and distribute the unique identity of users through steganographic images.
Xue \textit{et al.} \cite{ActiveGuard} exploit adversarial examples with specific conference as users' fingerprints to achieve users' fingerprints identification and active authorization control for DNN.

The emerging hardware architecture of DNN can also be considered as a hardware-level IP for device providers \cite{ChenFRZK19}. However, these smart devices may also be abused thus will threaten the IP of device providers.
Chen \textit{et al.} \cite{ChenFRZK19} proposed an attestation method for DNN devices (named DeepAttest), so as to provide hardware IP protection for DNN applications.
A device-specific fingerprint is designed and encoded in the weight of the DNN. The embedded fingerprint will then be extracted with the support of a Trusted Execution Environment (TEE) and used to verify whether the queried DNN has passed identity verification \cite{ChenFRZK19}.
DeepAttest ensures that only legitimate DNN programs can generate matching fingerprints, and allows it working on the target devices.
Chakraborty \textit{et al.} \cite{HAIPP} propose HANN, a hardware-assisted DNN IP protection approach. They obfuscate the weights of the model based on a secret key. The key is stored in a trusted hardware device. Users can only use the model if they can provide the trusted key device \cite{HAIPP}.
Cammarota \textit{et al.} \cite{CammarotaBR18} discussed the hardware mechanisms for machine learning IP protection in embedded terminal devices.
The above few hardware DNN copyright protection works focus on the IP protection for hardware DNN, instead of software DNN, and require hardware platform to support, such as TEE, hardware root of trust, which is costly.

\subsection{Target Models}
While most works focus on protecting the IP of classification models, Zhang \textit{et al.} \cite{zhang2020model} propose a watermarking approach for image processing models which are more complex. Specifically, a invisible watermark is embedded under black-box scenario by using spatial invisible watermarking schemes.
Quan \textit{et al.} \cite{ImageProcessing} exploit the overparameterization of models to embed watermarks for image processing tasks. Besides, they designed an auxiliary module to visually display the watermark information for verification.

The above watermarking methods require to control the training process and the training data. This can only be applied to the scenarios where a single entity trains the model locally \cite{WAFFLE}, but it is not suitable for distributed training scenarios, e.g., Federated Learning. Atli et. al. \cite{WAFFLE} propose a watermarking approach for Federated Learning scenarios. Each time the local model is aggregated to the global model, the model is re-trained to embed the backdoor/watermark. They also propose a watermark pattern generation method in which the images are generated with a random but class-specific pattern \cite{WAFFLE}.

\subsection{Function}
Most watermarking methods target at copyright verification, i.e., verifying the ownership of the model, which are passive copyright protection methods.
In commercial copyright management, active copyright management are required, i.e., active authorized control and users' identities management, including assigning a unique identity to each user, authenticating and managing the user's identity, and only authorized users are allowed to use the model.
In the above copyright verification or copyright management scenarios, robust watermarks are used.

From another aspect, watermarks can also be used to verify the integrity of the model, in which fragile or reversible watermarks are required.
Szentannai \textit{et al.} \cite{abs-1907-01650} proposed a fragile neural network to prevent model piracy.
They propose an approach for constructing an functional equivalent version of DNN, which has the same response and accuracy, but it is very sensitive to the modifications of the weights. The method generates a completely sensitive and fragile model, so that even a small weight changes of the model will greatly change the response of the model \cite{abs-1907-01650}.
Guan \textit{et al.} \cite{GuanFZZZY20} propose a reversible watermarking scheme for convolutional neural network to verify the integrity of the model, in which they generate a host sequence and embed the watermark through histogram shift. If the model is maliciously modified, the extracted watermark information will be completely different \cite{GuanFZZZY20}.

\section{Attacks on DNN IP Protection Works}
\label{sec4}
This paper systematically discusses the anti-attack ability of the DNN IP protection methods in the face of different levels of attackers.
As shown in Table \ref{tab:2},
we divide different types of attacks against the DNN IP protection methods into the following three levels (from weak to strong): (i) \textit{model modification}, refers to unintentional/common model modifications, including model fine-tuning, model pruning, model compression, model retraining, etc.; (ii) \textit{evasion attacks} and \textit{removal attacks}, refer to the passive attack methods, i.e., the attackers try to escape the watermark detection, including removal attacks, tampering, reverse-engineering attacks, etc.; (iii) \textit{active attacks}, refer to the active and strong attacks, including ambiguity attack, watermark/fingerprint collusion attack, watermark/fingerprint overwriting, query modification attack, etc.
In the following sections, we will discuss these three levels of attacks.

\begin{table*}[!htbp]
  \centering
  \caption{Attack Resistance of Existing DNN IP Protection Methods under Different Levels of Attacks}
    \begin{tabular}{|m{1.6cm}<{\centering}|m{2.5cm}<{\centering}|m{8cm}<{\centering}|m{2cm}<{\centering}|}
    \hline
    Attack level & Attack type & Attack method & Attack resistance \\
     \hline
    Level 1 & Model modifications & Model fine-tuning, Model pruning, Model compression, Retraining & \Checkmark \\
    \hline
    Level 2 & Evasion attacks, Removal attacks & Removal attacks, Tampering, Reverse-engineering attacks  & Partially \\
      \hline
    Level 3 & Active attacks & Ambiguity attack, Watermark detection, Watermark overwriting, Collusion attack, Query modification attack & \XSolid \\
     \hline
    \end{tabular}%
  \label{tab:2}%
\end{table*}%

\subsection{Level 1: Model Modifications}
After pirating a DNN model, the pirate often modifies or compresses the DNN model, and then deploys and uses it as a MLaaS to provide services. Therefore, most of the existing DNN IP protection works have evaluated the robustness against model modifications. Model modifications include:

1) Model fine-tuning \cite{UchidaNSS17}: Fine-tuning involves retraining the model to change model's parameters while maintaining the performance. Model fine-tuning can build many models based on existing models. Since the parameters carrying the watermark will be changed during the fine-tuning process, the embedded watermark should be robust to fine-tuning.

2) Model pruning or parameter pruning \cite{RouhaniCK19}: Model pruning is a common method for deploying DNN, especially on embedded devices. Honest users may use parameter pruning to reduce the memory and computational overhead of DNNs, while adversaries may use pruning to remove watermarks, such as using pruning methods to sparse weights in the watermarked DNN. Therefore, an effective watermarking technique should be able to resist parameter changes caused by parameter pruning.

3) Model compression \cite{UchidaNSS17}: Model compression can significantly reduce memory requirements and computational overhead, and is important for deploying DNN to embedded systems or mobile devices. Lossy compression will distort the model parameters, thus it is necessary to explore its impact on the detection rate of watermarks.

4) Model retraining \cite{NambaS19,abs-1910-12903}: A direct method of removing watermarks is to retrain the model with new instances, which can possibly remove the watermark or reduce the impact of the watermark.

\subsection{Level 2: Evasion Attacks and Removal Attacks}
Most works only evaluate the robustness of the watermark against unintentional \textit{model modifications}, but does not consider the security of the watermark when the attacker takes attacks.
In fact, DNN watermarks also face a variety of attacks. The common passive attacks are \textit{evasion attacks} \cite{abs-1809-00615} and \textit{removal attacks}.
Recently, there is a few works to consider the security of DNN watermarks against evasion attacks and removal attacks.

1) Removal attacks \cite{abs-1906-07745}: the attackers try to remove the watermark.

2) Tampering \cite{GuoP18}: The attacker knows that there is a watermark in the model. He attempts to tamper with the model to remove the IP owner's signature.

3) Reverse-engineering attacks \cite{FanNC19}: If pirates can obtain the original training dataset, they may reverse engineering the hidden parameters directly.

Shafieinejad \textit{et al.} \cite{abs-1906-07745} studied removal attacks towards backdoor-based watermarking schemes in DNN. Three attack methods, i.e., white-box, black-box, and property inference attack are used, respectively. They indicated that the adversary can remove the watermark by only relying on public data without accessing the training set, trigger set or model parameters \cite{abs-1906-07745}. They also propose a method to detect whether the model contains a watermark, indicating that the backdoor-based watermarking schemes are not secure enough to keep the watermark hidden \cite{abs-1906-07745}.
To remove the backdoor based DNN watermak, Wang \textit{et al.} \cite{GanRemoval} use GAN to detect and reverse the backdoor trigger in the model, and then fine-tune the model with the reversed trigger to remove the backdoor based watermark.
Hitaj and Mancini \cite{abs-1809-00615} evaluated the robustness and reliability of the DNN watermarking schemes, focusing on evasion attacks on DNN watermarks. They showed that even the watermark is difficult to remove, malicious attackers can still evade the owner's verification, so as to prevent model theft from being discovered \cite{abs-1809-00615}.
Chattopadhyay \textit{et al.} \cite{ChattopadhyayVC20} use GAN to generate samples for retraining which can obtain a model with similar performance while removing the watermark.

Earlier works demonstrated that simple fine-tuning could not remove the watermark, but recent studies have shown that improved fine-tuning can remove the watermark. Chen \textit{et al.} \cite{chen2019leveraging} leveraging unlabeled data to facilitate the fine-tuning based watermark removal. By using a pre-trained DNN to label the unlabeled data, the number of labeled data required by the attacker for watermark removal can be reduced significantly.
Chen \textit{et al.} \cite{REFIT} propose an fine-tuning based watermark removal scheme by using carefully designed learning rate schedule. Specifically, they incorporate two techniques into the scheme, elastic weight consolidation and unlabeled data augmentation.
Liu \textit{et al.} \cite{LiuRemoving} propose a framework to remove backdoor-based watermarks with limited data, named WILD. Specifically, a data augmentation method is proposed to imitate the behavior of the backdoor triggers.

Aiken \textit{et al.} \cite{NNLaundering} propose a DNN laundering scheme to remove backdoor-based watermarks. The approach consists of three steps: watermark recovery, watermarked neurons resetting, and retraining.
Guo \textit{et al.} \cite{HiddenVulnerability} propose a watermark removal attack without prior knowledge. They used a preprocessing operation, which added perturbation and transformations to the input, making the watermark trigger invalid. Then, they use fine-tuning with unlabeled data to improve the performance of the model \cite{HiddenVulnerability}.

\subsection{Level 3: Active Attacks}
We believe that it is not enough to only consider the above-mentioned robustness to model modifications and security to evasion/removal attacks. It should also evaluate the anti-attack performance of the model under active and strong attacks, as follows:

1) Ambiguity attack \cite{FanNC19}: Ambiguity attack aims to doubt the ownership verification by forging an additional watermark on the DNN model.
For example, in the case of DNN authorization control, the adversary's purpose is to deceive the DNN that the input comes from an authorized user.
In the context of traditional digital watermarking techniques, studies have shown that unless an irreversible watermarking scheme is adopted, a robust watermark may not necessarily verify ownership \cite{FanNC19}.

2) Watermark detection: recently, a few works have been proposed focusing on how to detect watermarks (such as backdoors) in order to take further attacks.

3) Watermark overwriting \cite{RouhaniCK19, abs-1904-00344, ChenRFZK19}: If an attacker knows the watermark embedding method in the model (but does not know the owner's private watermark information), he may want to overwrite the original watermark by embedding a new watermark in the deep learning model, so as to destroy the original watermark or make the original watermark unreadable.

4) Watermark/fingerprint collusion attack \cite{ChenRFZK19}: A group of users with the same host DNN and different fingerprints may conduct a collusion attack to build a functional model that can prevent the copyright owner from verifying the ownership \cite{ChenRFZK19}.

5) Query modification attack \cite{NambaS19}: The pirates modify the query to invalidate the watermark verification process. Specifically, after the pirate has deployed the pirated MLaaS service, the pirate will actively detect whether a query is the watermark verification query from the IP owner, thereby modifying or shielding the query to make the watermark verification process fail \cite{NambaS19}.

Wang and Kerschbaum \cite{WangK19} demonstrate that the standard deviation of the weights increases as the length of the embedded watermark increases. Therefore, by observing the standard deviation of the weights, an attacker can not only detect the watermark, but can also obtain the length of the watermark, which can be used for overwriting attacks \cite{WangK19}.

Chen \textit{et al.} \cite{ChenRFZK19} showed that in a large model distribution system, multiple users can use their respective watermarking models to collaboratively build a watermark-free model, which can achieve comparable accuracy as the original model. This attack is called the fingerprint collusion attack.
To this end, they proposed a fingerprint based digital rights management framework for deep learning models, called DeepMarks \cite{ChenRFZK19}.
DeepMarks enables owners to verify the IP information and user's unique identity, and can resist fingerprint collusion attacks.
Specifically, DeepMarks designs unique fingerprints for users and encodes each fingerprint in the probability density function of weights during DNN retraining with fingerprint-specific regularization losses \cite{ChenRFZK19}.

Wang and Kerschbaum \cite{wang2019robust} use another DNN to launch the property inference attack to detect the DNN watermarks. Then, they design a watermark scheme to resist the above detection framework. Specifically, they build an adversarial training structure similar to GAN where the watermark embedding and the detection network are the generator and the discriminator, respectively \cite{wang2019robust}.

Yang \textit{et al.} \cite{Effectiveness} demonstrate that distillation can be used as a strong attack method to remove DNN watermarks. The reason is that the watermark embedding has nothing to do with the main task. The distillation will remove the redundant information, including the watermark. To resist distillation attacks, they propose the ingrain watermark method, in which the embedding of watermark is inseparable from the main task \cite{Effectiveness}.

Namba and Sakuma \cite{NambaS19} proposed \textit{query modification} attack which worked as follows \cite{NambaS19}: (i) Key instance detection: when a query arrives, the pirate will use an autoencoder to detect whether the query sent by someone is a key instance for watermark verification. (ii) Query modification: if the query is determined as a watermark key instance, it will be modified (using an autoencoder to remove the logo from the image) to hinder the watermark verification process.
Otherwise, no changes will be made to the query \cite{NambaS19}. To overcome this attack, they propose a robust watermarking scheme with exponential weighting \cite{NambaS19}. Only parameters with large absolute values are used for prediction. As a result, the model can tolerate query modification attacks.

\section{Evaluation Suggestions for DNN IP Protection Methods}
\label{sec5}
Most of the evaluations of the existing DNN IP protection works only focus on the functional metrics \cite{UchidaNSS17, NagaiUSS18, RouhaniCK19, abs-1811-03713, abs-1904-00344, ChenRFZK19, ChenFRZK19, GuoP18} of the DNN watermark. We suggest to build the evaluation method for DNN IP protection methods from the following aspects:

1) Systematic evaluation method. We suggest to: (i) evaluate the performance of DNN IP protection methods under different levels of attacks, so as to clearly reveal the performance of DNN IP protection methods under different levels of attacks; (ii) establish comprehensive metrics to evaluate the performances of DNN IP protection methods; (iii) characterize the requirements of effective watermark/fingerprint authentication methods in the context of deep learning \cite{RouhaniCK19, abs-1904-00344, ChenRFZK19}. Such metrics can provide references for model designers and facilitate systematic comparison of existing and future DNN IP protection methods.

2) Basic functional metrics \cite{UchidaNSS17,NagaiUSS18,RouhaniCK19,abs-1811-03713,abs-1904-00344,ChenRFZK19,ChenFRZK19,GuoP18}, including: \textit{fidelity}, \textit{robustnesss}, \textit{functionality}, \textit{capacity}, \textit{efficiency}, \textit{reliability}, \textit{generality}, \textit{uniqueness}, \textit{indistinguishability}, and \textit{scalability}, as shown in Table \ref{tab:3}.

3) Attack-resistance metrics, including: \textit{security} \cite{UchidaNSS17,RouhaniCK19}, \textit{unremovability} \cite{AdiBCPK18,abs-1906-00830}, \textit{unforgeability} \cite{AdiBCPK18}, \textit{non-ownership piracy} \cite{AdiBCPK18,abs-1906-00830}, \textit{ownership piracy} \cite{AdiBCPK18,abs-1906-00830}, \textit{verifiability} \cite{AdiBCPK18}, \textit{collusion resistance} \cite{abs-1906-00830}, and \textit{non-invertible} \cite{FanNC19}, as shown in Table \ref{tab:3}.

4) Customized metrics for different application scenarios. As discussed in Section 2, there are different application scenarios for IP protection methods. Accordingly, different types of watermarks are needed, e.g., robust watermark, fragile watermark. The metrics for robust watermarks and fragile watermarks are different.

\begin{table*}[!htbp]
\centering
 \caption{Metrics for DNN IP Protection Methods}
   \resizebox{\textwidth}{!}{
    \begin{tabular}{|p{3.5cm}<{\centering}|m{11cm}|}
    \hline
    \multicolumn{2}{|p{36.19em}|}{\textbf{\textcircled{1} Basic functional metrics} \cite{UchidaNSS17,NagaiUSS18,RouhaniCK19,abs-1811-03713,abs-1904-00344,ChenRFZK19,ChenFRZK19,GuoP18}} \\
    \hline
    Fidelity & The function and performance of the model should not be affected by embedding the watermark. \\
    \hline
    Robustnesss & The watermarking method should be able to resist model modifications, such as compression, pruning, fine-tuning. \\
    \hline
    Functionality & It can support ownership verification, can use watermark to uniquely identify the model, and clearly associate the model with the IP owner's identity. \\
    \hline
    Capacity & The amount of information that the watermarking method can embed. \\
    \hline
    Efficiency & The watermark embedding and extraction processes should be fast with negligible computational and communication overhead. \\
    \hline
    Reliability & The watermarking method should ensure the least false negatives (FN) and false positives (FP), and the watermark key can be used to effectively identify the watermarked model. \\
    \hline
    Generality & The watermarking scheme can be applicable to white-box and black-box scenarios, various datasets and architectures, and various computing platforms. \\
    \hline
    Uniqueness & The target model's watermark should be unique. Further, each user's identity should also be unique. \\
    \hline
    Indistinguishability & The attacker cannot distinguish the wrong prediction from the correct model prediction. \\
    \hline
    Scalability & The watermark verification technique should be able to verify DNNs of different sizes. \\
    \hline
    \multicolumn{2}{|p{36.19em}|}{\textbf{\textcircled{2} Attack-resistance Metrics}} \\
    \hline
    Security \cite{UchidaNSS17,RouhaniCK19} & The watermark should be secret and should not be detected, read, modified or removed by unauthorized entities. \\
    \hline
    Unremovability \cite{AdiBCPK18,abs-1906-00830} & Even if the attacker knows the existence of the watermark and the embedding scheme, the watermark cannot be removed. \\
    \hline
    Unforgeability \cite{AdiBCPK18} & The adversary cannot convince the third party that he owns the copyright of the model. \\
    \hline
    Non-ownership piracy \cite{AdiBCPK18,abs-1906-00830} & The attacker is unable to generate a watermark for the watermarked model, so as to challenge the copyright owner's ownership. \\
    \hline
    Ownership piracy \cite{AdiBCPK18,abs-1906-00830} & The attacker tries to embed his watermark into a previously watermarked model. In this case, at least the old watermark must be preserved. A stronger requirement is that the new watermark can be distinguished from the old one or can be easily removed without knowing it. \\
    \hline
    Verifiability \cite{AdiBCPK18} & A watermarking scheme that can be verified using the verification process is called private verification. As long as the third party is honest and does not reveal the key, the owner can use verification to convince the third party that he has the ownership. \\
    \hline
    Collusion resistance \cite{abs-1906-00830} & Even if multiple API clients conspire to launch an extraction attack, the watermark is still undeletable and indistinguishable. \\
    \hline
    Non-invertible \cite{FanNC19} & Fan \textit{et al.} \cite{FanNC19} indicated that the ambiguity attack is mainly due to the inherent weakness (reversibility) of the watermark-based method. Therefore, designing an irreversible watermark verification scheme plays an indispensable role in eliminating ambiguity attacks. \\
    \hline
    \end{tabular}}
    \label{tab:3}\\
\end{table*}

\section{Challenges and Future Works}
\label{sec6}
\subsection{Challenges}
Deep learning models are deployed on white-box or black-box platforms, and have complex structures and massive parameters. Developing copyright protection method for deep learning models that can be used in commercial applications is a difficult problem.
In summary, the existing DNN IP protection works face the following challenges:

1) Most of the existing DNN IP protection works are passive verification methods, i.e., verifying the copyright of the DNN model after the piracy occurs. Such post-verification methods cannot actively prevent the occurrence of the piracy.

2) Most of the existing DNN IP protection methods only focus on verifying the ownership of the model, but do not authenticate and manage the users' identities, thus are not suitable for commercial applications. Besides, these methods are vulnerable to attacks launched by dishonest users, such as collusion attacks.

3) Most of the attack-resistance evaluated in the existing DNN IP protection works are model modifications, without taking into account of the active and strong attacks of the adversary, such as query modification attacks, collusion attacks, and ambiguity attacks, etc. If pirates take active attacks, there are some novel and powerful attack methods that can defeat most of the existing DNN IP protection methods. Potential active attacks and countermeasures are open issues.

4) Lack of systematic evaluation method. As an emerging research direction, DNN IP protection is still in its infancy. It lacks a systematic evaluation method, especially lack of the evaluation of DNN IP protection methods in the face of pirates taking active attacks.

\subsection{Future Works}
The potential future directions on DNN IP protection are as follows:

(1) Active attacks and corresponding countermeasures.
The attack-resistance of existing DNN IP protection works are mostly focusing on model modifications (a small amount of works consider watermark removal attacks), without evaluating the attack-resistance under active attacks.
If the pirates take active attacks, such as collusion attacks and query modification attacks, the existing DNN IP protection methods may fail. The difficulty lies in how to resist the active and powerful attacks under the premise of ensuring reliable watermark extraction, and not affecting model's performance in deep learning scenarios.
New types of attacks and corresponding defenses are endless competitive games.

(2) Fragile/reversible watermarking methods. Most of the existing works focus on robust watermarking, but in special scenarios, such as integrity verification of the model, fragile watermarking is needed. This is a topic that has received little attention at present.

(3) Active authorization management mechanism for DNN models.
Most of the existing methods are passive verification methods which can only verify the copyright after piracy occurs. An active DNN IP protection method which can lock the model, actively prevent piracy (realizing \textit{copyright protection}), and manage the users' identities (realizing \textit{copyright management}), is needed. The difficulty lies in how to control the functions and performance of the deep learning model differently according to different users.

(4) Management of users' identities for DNN models. Most of the existing methods construct watermarks on the model to verify the ownership of the model, but cannot achieve commercial copyright management. Studying the management of users' identities for DNN can provide mature solutions for commercial applications. The difficulties lie in: (i) how to design unique identity for each user; (ii) how to authenticate and track users' identities; (iii) how to differentially control the performance of the model according to the users' identities;
(iv) how to make DNN be able to distinguish between authorized users and unauthorized users;
(v) how to make DNN be able to distinguish different authorized users;
(vi) and how to overcome ambiguity attacks and collusion attacks launched by dishonest users.

(5) Fast and efficient watermark verification algorithm. In the existing watermarking methods, the watermark extraction or verification process is inefficient.
In the context of software and image, there have been effective hash algorithms that can be used for fast search and verification \cite{xue2019ssl}. However, in the context of deep learning, fast, efficient and large-scale search, watermark extraction and verification methods are still lacking, thus cannot meet the requirements of practical commercial copyright management.

(6) Not only IP protection for models, but also IP protection for data. Most of existing works focus on protecting the IP of the models. However, in deep learning scenarios, in addition to models, the data, including training data and output data, are also valuable and can be regarded as IP. For example, the private training dataset collected and annotated by the company can also be regarded as the company's IP. The paintings output by the deep learning model have certain values thus also require effective IP protection. In works \cite{WatermarkedImages, zhang2020model}, the output images of an watermarked model will also contain the watermark, which can be applied to protect the IP of the output data.

\section{Conclusions}
\label{sec7}
DNN copyright protection is a valuable potential research topic which has received more and more concerns in recent years.
This paper presents a survey on existing DNN IP protection methods, focusing on the challenges these methods face, whether these methods can provide proactive protection, and different levels of attacks.
Besides, the first taxonomy on DNN IP protection works is proposed, and the systematic evaluation suggestions for DNN IP protection methods is presented.
DNN IP protection is still in its infancy. The challenges include: most of the existing works focus on copyright verification and cannot support copyright management; most of the existing works are passive verification methods rather than active control methods; unable to resist active and powerful attacks, etc.
A DNN IP protection method which combines copyright verification and copyright management, using active authorization control, and can resist active attacks, is a potential future direction.
This paper can hopefully provide a reference for the taxonomy, comparison, evaluation and development of DNN IP protection methods.

\bibliographystyle{IEEEtran}
\bibliography{ref}

\end{document}